\documentclass[a4paper, 10pt, conference]{IEEEtran}
\IEEEoverridecommandlockouts
\usepackage{cite}
\usepackage{amsmath,amssymb,amsfonts}
\usepackage{algorithmic}
\usepackage{graphicx}
\usepackage{textcomp}
\usepackage{amsmath,amsfonts}
\usepackage{algorithmic}
\usepackage{algorithm}
\usepackage{array}
\usepackage{subcaption}

\usepackage{geometry}
\usepackage{stfloats}
\usepackage{url}
\usepackage{verbatim}
\usepackage{graphicx}
\usepackage{cite}
\usepackage{tikz}
\newcommand*\circled[1]{\tikz[baseline=(char.base)]{
            \node[shape=circle,draw,inner sep=1pt, font=\scriptsize, anchor=mid] (char) {#1};}}

\usepackage{xcolor}
\usepackage[T1]{fontenc}        
\usepackage{mathptmx}           
\usepackage{newtxtext,newtxmath}  

\def\BibTeX{{\rm B\kern-.05em{\sc i\kern-.025em b}\kern-.08em
    T\kern-.1667em\lower.7ex\hbox{E}\kern-.125emX}}

\makeatletter
\newcommand{\linebreakand}{%
  \end{@IEEEauthorhalign}
  \hfill\mbox{}\par
  \mbox{}\hfill\begin{@IEEEauthorhalign}
}
\makeatother
\begin{document}

\title{WPTrack: A Wi-Fi and Pressure Insole Fusion System for Single Target Tracking\\
}

\author{\IEEEauthorblockN{1\textsuperscript{st} Wei Guo}
\IEEEauthorblockA{\textit{Industrial and Systems Engineering} \\
\textit{The Hong Kong Polytechnic University}\\
Hong Kong, China \\
guo-wei.guo@polyu.edu.hk}
\and
\IEEEauthorblockN{2\textsuperscript{nd} Shunsei Yamagishi}
\IEEEauthorblockA{\textit{Computer Science and Engineering} \\
\textit{University of Aizu}\\
Aizuwakamatsu, Japan \\
d8271107@u-aizu.ac.jp}
\and
\IEEEauthorblockN{3\textsuperscript{rd} Lei Jing}
\IEEEauthorblockA{\textit{Computer Science and Engineering } \\
\textit{ University of Aizu}\\
Aizuwakamatsu, Japan \\
leijing@u-aizu.ac.jp}

\linebreakand

}

\maketitle

\begin{abstract}
As the Internet of Things (IoT) continues to evolve, indoor location has become a critical element for enabling smart homes, behavioral monitoring, and elderly care. Existing WiFi-based human tracking solutions typically require specialized equipment or multiple Wi-Fi links, a limitation in most indoor settings where only a single pair of Wi-Fi device is usually available. However, despite efforts to implement human tracking using one Wi-Fi link, significant challenges remain, such as difficulties in acquiring initial positions and blind spots in DFS estimation of tangent direction. To address these challenges, this paper proposes WPTrack, the first \textbf{\underline{W}}i-Fi and \textbf{\underline{P}}ressure Insoles Fusion System for Single Target \textbf{\underline{Track}}ing. WPTrack collects Channel State Information (CSI) from a single Wi-Fi link and pressure data from 90 insole sensors. The phase difference and Doppler velocity are computed from the CSI, while the pressure sensor data is used to calculate walking velocity. Then, we propose the CSI-pressure fusion model, integrating CSI and pressure data to accurately determine initial positions and facilitate precise human tracking. The simulation results show that the initial position localization accuracy ranges from 0.02 cm to 42.55 cm. The trajectory tracking results obtained from experimental data collected in a real-world environment closely align with the actual trajectory.
\end{abstract}


\section{Introduction}
Indoor human tracking is essential in numerous smart applications, such as smart home systems, elderly care, and large-scale indoor scene navigation. Although satellite navigation has achieved remarkable accuracy in outdoor environments, it cannot be used indoors due to buildings blocking the satellite signals. Consequently, the challenge of implementing indoor positioning has sparked interest among researchers, leading to the proposal of various solutions to address indoor navigation issues. 

Camera-based systems\cite{Xiao2019HumanTF} for human localization and tracking offers a direct method. However, issues such as visual blind spots, privacy concerns, and the need for adequate lighting limit their popularity. Acoustic-based solution\cite{Xu2019AcousticIDGH} does not have these drawbacks, but it suffers from a limited coverage range. IMU-based method\cite{Yuan2014LocalizationAV} is not disturbed by external environmental factors, yet it suffers from accumulated tracking errors over time. To obtain the absolute location and improve performance, it needs to be integrated with other sensors, such as cameras\cite{Poulose2019HybridIL} and pressure insoles\cite{Chen2019SmartII}. In light of the ubiquitous integration of radio frequency (RF) technologies in daily life, researchers are exploring the use of RF signals for human tracking, which includes technologies like Radio Frequency Identification (RFID)\cite{Zhang2023RealTimeAA}, Ultra-Wideband (UWB)\cite{ridolfi2021self}, millimeter wave (mmWave)\cite{Zhang2023ASO}, Bluetooth\cite{Oosterlinck2017BluetoothTO} and Wi-Fi\cite{Liu2020SurveyOW}. Among these technologies, Wi-Fi is particularly noteworthy for advantages such as no need for additional deployment, extensive coverage and cost-effectiveness. 

Wi-Fi signals enable the acquisition of the Received Signal Strength Indicator (RSSI) and Channel State Information (CSI). RSSI is commonly used for fingerprint localization, and researchers have used deep learning algorithms to try to reduce the labor involved in taking fingerprints \cite{Vishwakarma2023IndoorGNNAG}. CSI is a finer-grained information than RSSI, which does not produce large fluctuations in the same environment and is commonly used to compute Angle of Arrival (AoA), Time of Flight (ToF), and Doppler Frequency Shift (DFS) for target location tracking. SpotFi \cite{Kotaru2015SpotFiDL} enhances localization technology by integrating super-resolution algorithms for precise AoA computation from multipath components with just three antennas, achieving decimeter-level precision in active localization. WiTraj \cite{Wu2023WiTrajRI}, DFT-JVAE \cite{Zhang2019DeviceFreeTV}, IndoTrack \cite{Li2017IndoTrackDI} propose innovative algorithms to calculate parameters such as AoA, ToF, and Doppler velocity through human-reflected signals, enabling passive localization. However, these works are predicated on the existence of multiple Wi-Fi links, an impracticality in real-world indoor environments due to link scarcity. Consequently, Widar2.0 \cite{Qian2018Widar20PH} pioneers a passive trajectory tracking technique using a solitary Wi-Fi link. Utilizing the SAGE algorithm, it calculates parameters including amplitude, AoA, ToF, and DFS, while graph-based path matching facilitates the estimation of walking routes. Subsequently, works such as WiDFS \cite{Wang2021SingleTargetRP}, WiSen \cite{Jin2022WiSenZP}, and WiMT \cite{Guo2023TowardLP} have also introduced systems for passive trajectory tracking using a single Wi-Fi link, attempting to derive more accurate parameters from limited information for trajectory estimation. However, in single-link systems, where analysis of target-reflected signals yields AoA, DFS, and ToF information, the low time resolution of Wi-Fi makes ToF estimation challenging. Furthermore, AoA and DFS estimations often encounter dead zones; for instance, if the target moves in the direction of the current AoA, there will be no change in AoA, or if it moves parallel to the Line of Sight (LoS), making path variations subtle and thus hindering DFS acquisition. To address the challenges encountered in single-link target tracking, we propose WPTrack, the first \textbf{\underline{W}}i-Fi and \textbf{\underline{P}}ressure Insoles Fusion System for Single Target \textbf{\underline{Track}}ing. WPTrack collects CSI from a single Wi-Fi link pair and pressure data from 90 insole sensors. The phase difference and Doppler velocity are computed from the CSI, while the pressure sensor data is used to calculate walking velocity. Then, we propose the CSI-pressure fusion model, integrating CSI and pressure data to accurately determine initial positions and facilitate precise human tracking. The key insight of WPTrack is that within the sensing area, no matter where the target moves, the target will be on the Fresnel Zone with Tx and Rx as the focus in space. The line connecting the target to Tx always forms an ellipse intersecting at the target's location. 

To realize WPTrack, our main contributions are as follows:

\begin{itemize}
    \item We propose the first work to integrate Wi-Fi and pressure insoles for indoor localization and human tracking. 
    \item WPTrack proposes a CSI-pressure fusion model that addresses Wi-Fi localization challenges such as tangential tracking blind spots with a pair of Wi-Fi devices and complex initial estimates. It also solves issues of pressure insole trajectory estimates, eliminating the need for complex IMUs and enabling absolute positioning.

\end{itemize}


\section{Preliminary}

\subsection{Channel State Information}

In a narrow-band flat fading channel, the Wi-Fi OFDM system viewed in the frequency domain can be defined as\cite{Guo2023TowardLP}:
\begin{equation}
\vec{Y}= H \vec{X} + \vec{N}
\label{eq1}
\end{equation}
where $\vec{Y}$ and $\vec{X}$ represent the received and transmitted signal vectors, respectively. $H$ denotes the Channel Frequency Response (CFR) and $\vec{N}$ is the Additive White Gaussian Noise (AWGN). The CFR in time and frequency as amplitude and phase in the format of CSI, is a superposition of signals from multipath propagation. Hence, the CSI can be represented as\cite{Guo2023TowardLP}:

\begin{equation}
H( f_s ,t)= \displaystyle\sum\limits_{k=1}^{L} A_{s,k} e^{-j2 \pi f_n\tau_k(t)} 
\label{eq2}
\end{equation}
where $f_s$ is the carrier frequency of the $s^{th}$ subcarrier, $s$ is the index of the \textcolor{black}{OFDM} subcarrier, $s\in[1,30]$. $L$ is the number of paths, $A_{s,k}$ denotes the amplitude and $\tau_k(t)$ represents the propagation time of the $k^{th}$ path. Moreover, the phase of CSI at carrier frequency $f_s$ propagating along the $k^{th}$ path can be written as $\varphi_{s,k}=2 \pi f_s\tau_k(t)$, the $\tau_k(t)$ represents the signal propagation delay\cite{Guo2023TowardLP}. 

\subsection{Pressure Insole}

\begin{figure}[htbp]
    \centering
    \includegraphics[width=0.21\textwidth]{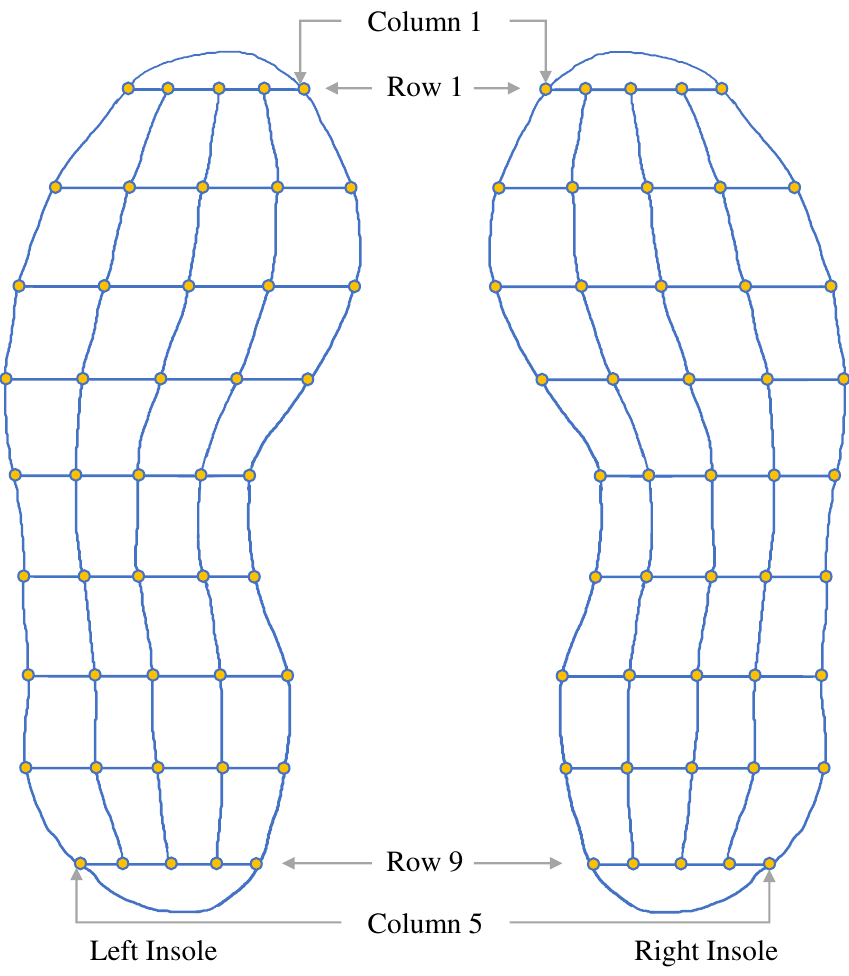} 
    \caption{Pressure insoles.} 
    \label{fig:pressure insole} 
\end{figure}

We employed pressure insoles fabricated by PIFall \cite{Guo2024PIFallAP} to capture the plantar pressure of the subject. Each insole is intricately stitched with resistive films and conductive threads, forming 45 discrete pressure sensors. As depicted in Fig. \ref{fig:pressure insole}, the sensors are arranged into a 9x5 matrix and are non-uniformly distributed across the insole. Upon the application of pressure by the foot, the resistance of the sensors decreases, leading to an increase in voltage, and vice versa.

\section{Method}
\begin{figure}[htbp]
    \centering
    \includegraphics[width=0.44\textwidth]{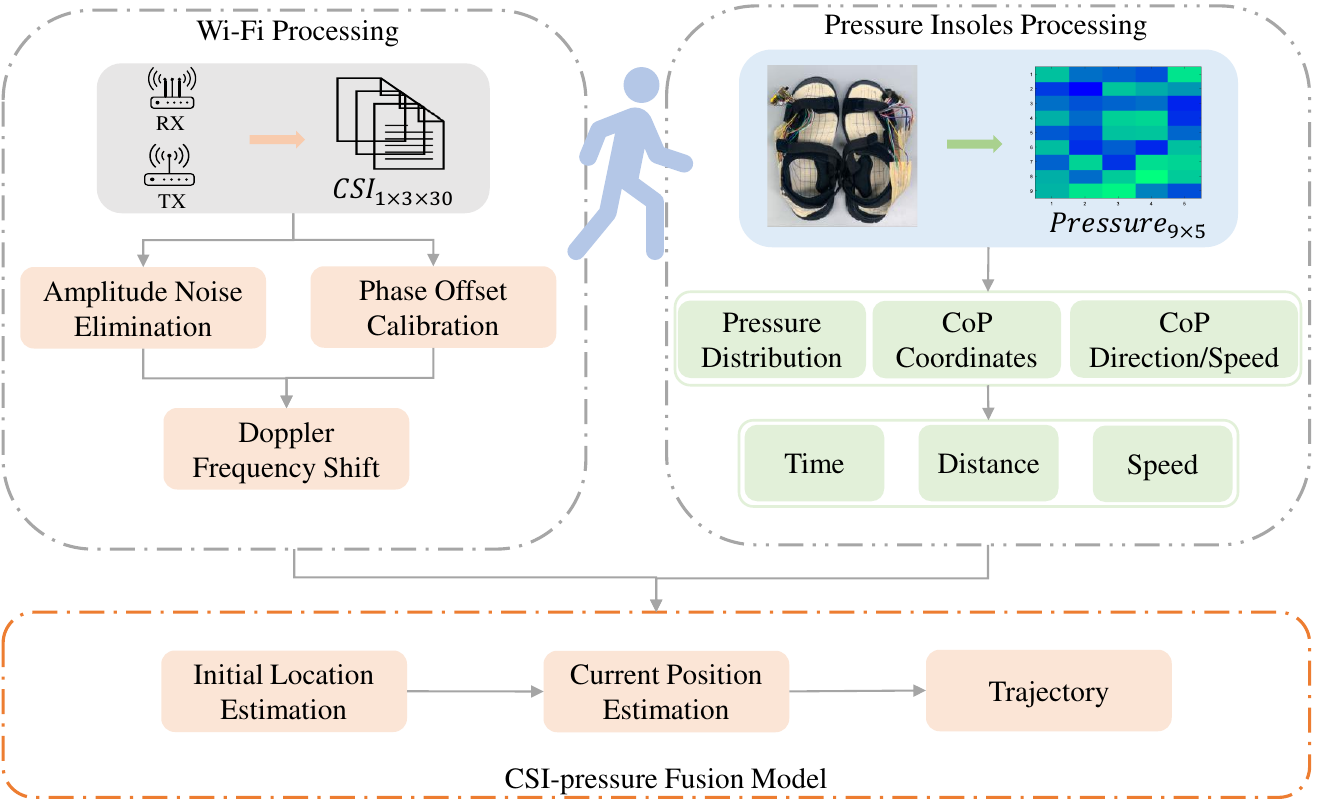} 
    \caption{System Overview.} 
    \label{fig:system_overview} 
\end{figure}

The system overview is illustrated in Fig. \ref{fig:system_overview}. Initially, CSI and pressure data are collected independently from Wi-Fi signals and pressure insoles, respectively, followed by a preprocessing phase. The CSI is utilized to calculate the Doppler velocity and AoA, while data from the pressure insoles are employed to ascertain the walking duration, distance, and velocity of the target. Subsequently, these parameters are fed into the CSI-pressure model to estimate the target's coordinates and heading angle. Ultimately, the aggregation of these coordinates forms a trajectory for target tracking.

\subsection{Wi-Fi Processing}

\subsubsection{De-noising}
We use Intel 5300 NIC to receive CSI data, TX is equipped with one antenna, RX is equipped with three antennas. Therefore, the dimension of the CSI data matrix is $1\times3\times30$. Due to the imperfections in hardware devices, the phase of CSI contains phase noise introduced by carrier frequency offset (CFO), sampling frequency offset (SFO), and packet detection delay (PDD), while the amplitude of CSI is polluted by environmental noise. We employ the Savitzky-Golay filter and linear transform method \cite{wu2015phaseu,Wang2017RTFallAR,Nan2020LeveragingTT} to remove noise from the amplitude and phase, respectively.

\subsubsection{AoA Calculation}
AoA estimation methods have been studied in many works, such as \cite{Li2016DynamicMUSICAD, Qian2018Widar20PH}, and we directly choose the existing methods to calculate AoA.

\subsubsection{Doppler Frequency Shift Calculation}

\begin{figure}[htbp]
    \centering

    \begin{subfigure}[b]{0.45\linewidth}
        \includegraphics[width=\linewidth]{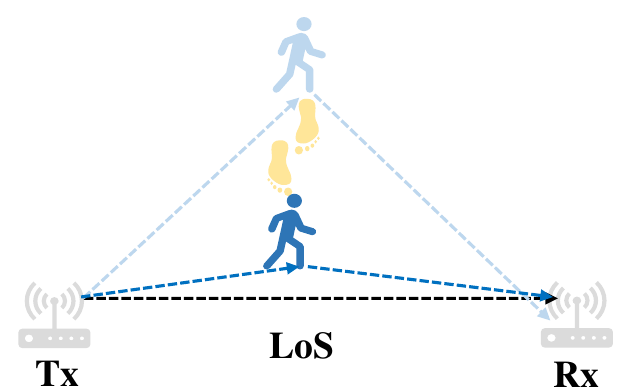}
        \caption{normal walk}
        \label{fig:sub3}
    \end{subfigure}
    \quad  
    \begin{subfigure}[b]{0.45\linewidth}
        \includegraphics[width=\linewidth]{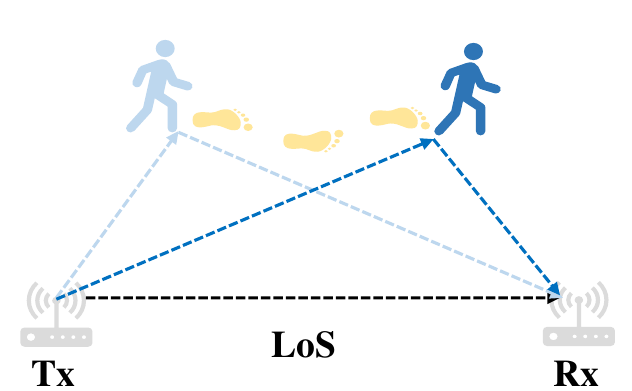}
        \caption{tangent walk}
        \label{fig:sub4}
    \end{subfigure}

    \vspace{1em}  

    \begin{subfigure}[b]{0.45\linewidth}
        \includegraphics[width=\linewidth]{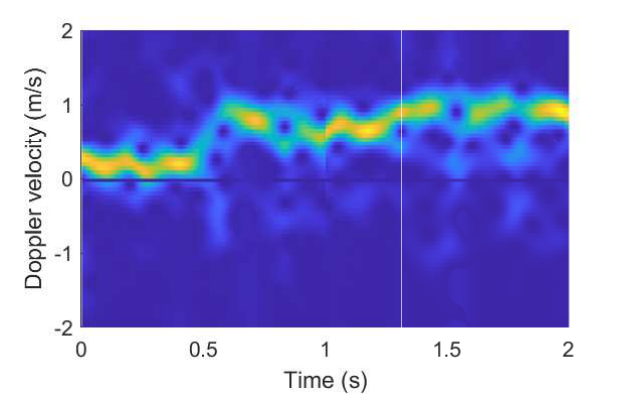}
        \caption{normal doppler}
        \label{fig:sub1}
    \end{subfigure}
    \quad  
    \begin{subfigure}[b]{0.45\linewidth}
        \includegraphics[width=\linewidth]{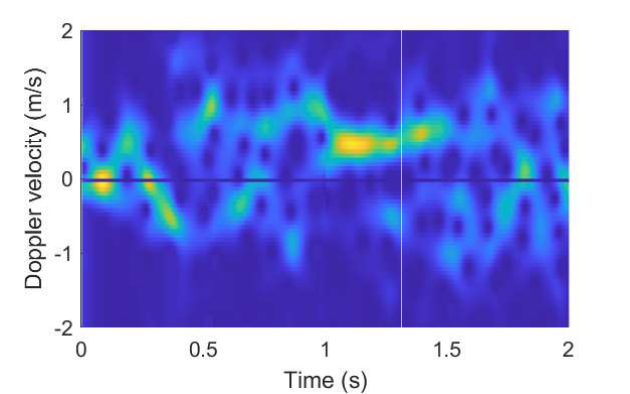}
        \caption{tangent doppler}
        \label{fig:sub2}
    \end{subfigure}

    \caption{Doppler of two walking directions}
    \label{fig:normal and tangent}
\end{figure}

We employ the method outlined in study \cite{Guo2023TowardLP} for the computation of Doppler velocity. The target's movement along the directions illustrated in Fig. \ref{fig:normal and tangent} (a) and (b) results in the Doppler velocities depicted in Fig. \ref{fig:normal and tangent} (c) and (d), respectively. It is evident from the figures that despite identical walking distances and speeds, the Doppler velocities differ. The normal direction yields superior results compared to the tangential direction, as movement along the tangential direction does not significantly alter the path of the reflected signal, leading to tracking blind spots.

\subsection{Pressure Insole Processing}
\begin{figure}[htbp]
    \centering
    \includegraphics[width=0.45\textwidth]{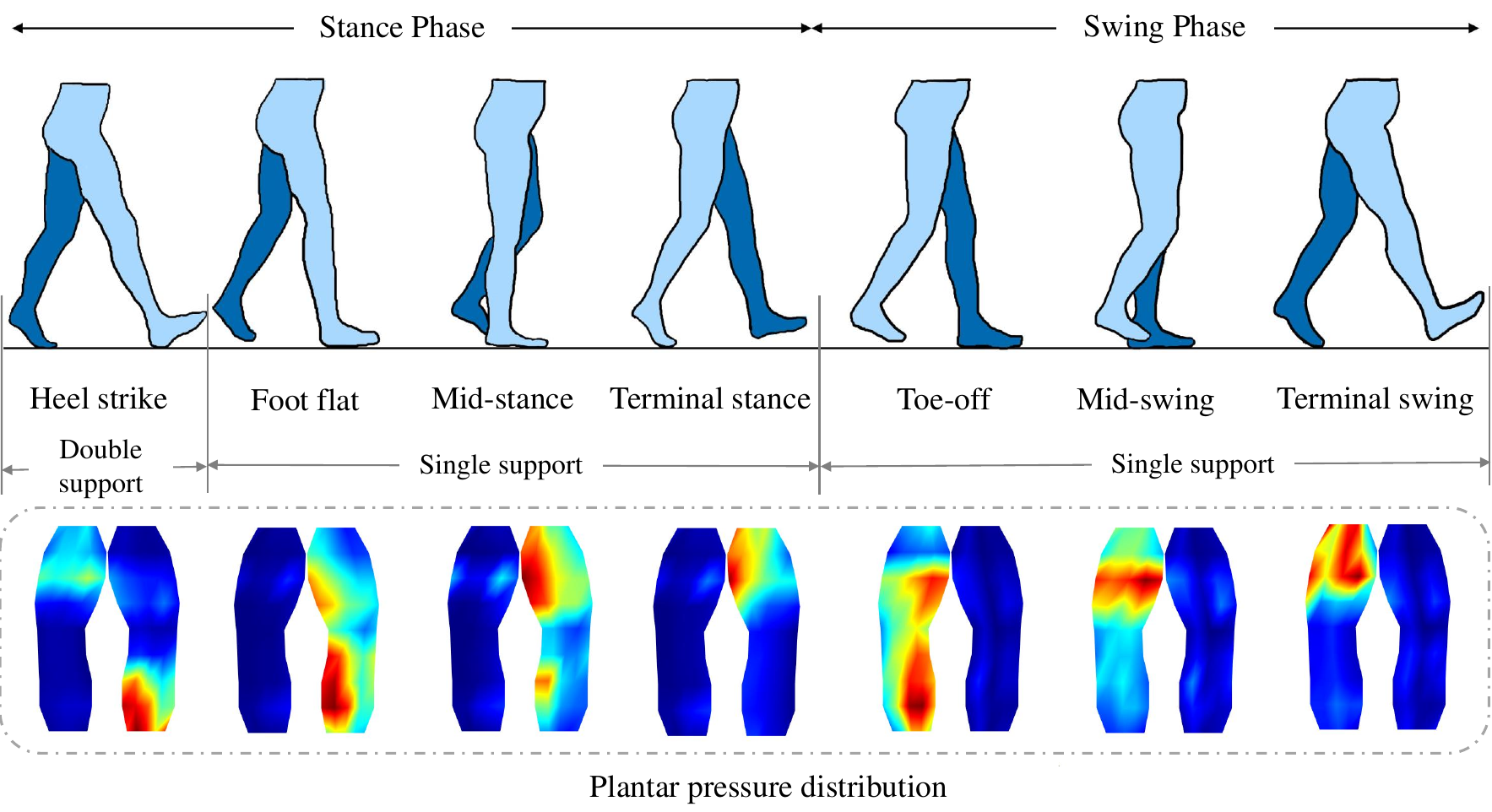} 
    \caption{Gait cycle and planter pressure distribution.} 
    \label{fig:gait cycle pressire dis} 
\end{figure}

By monitoring fluctuations in plantar pressure, pressure insoles effectively map out the dynamics of the gait cycle. Fig. \ref{fig:gait cycle pressire dis} depicts the plantar pressure values of both feet at seven instances throughout a complete gait cycle, where blue areas indicate lower pressure values and red areas represent higher pressure values. From the figure, it is observable that the pressure insoles accurately correspond to each phase of the gait cycle. 

\subsubsection{Time synchronization}

\begin{figure}[ht]
  \centering
  \begin{subfigure}[b]{0.46\linewidth} 
    \includegraphics[width=\linewidth]{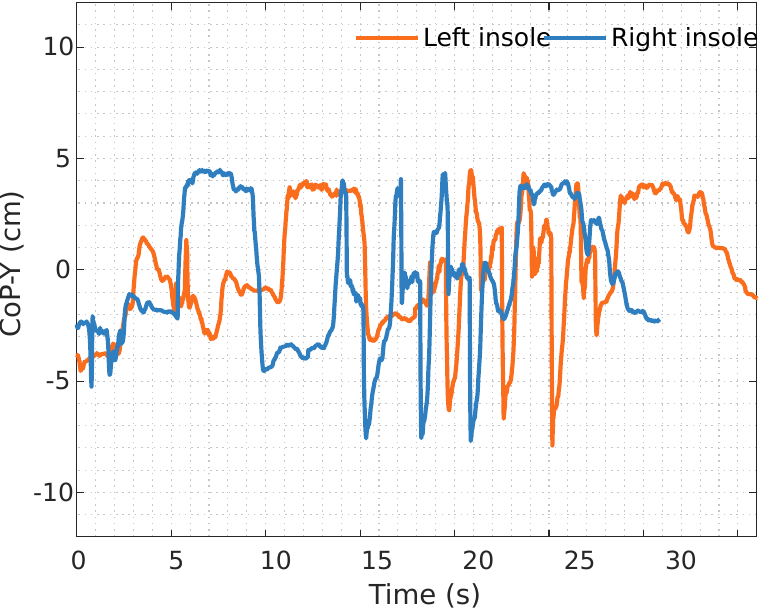}
    \caption{Raw data}
    \label{fig_align:sub1}
  \end{subfigure}%
  \hspace{3mm} 
  \begin{subfigure}[b]{0.47\linewidth} 
    \includegraphics[width=\linewidth]{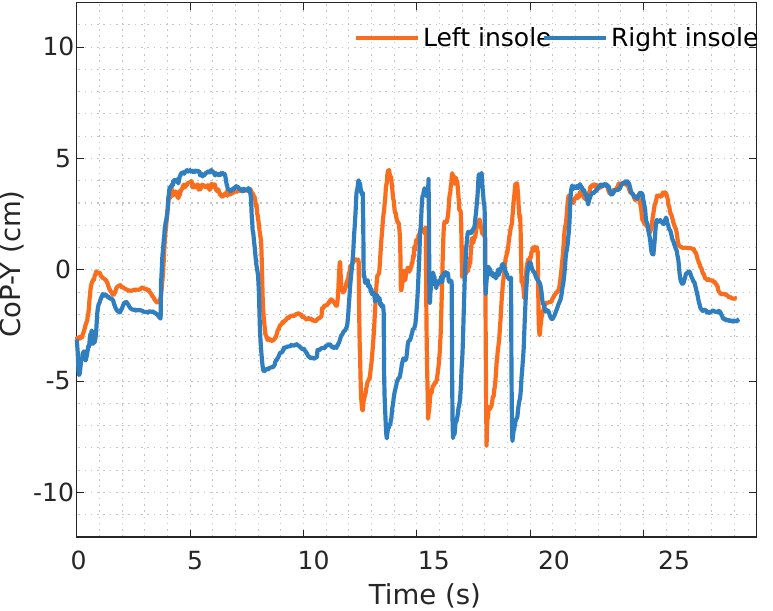}
    \caption{Alignment data}
    \label{fig_align:sub2}
  \end{subfigure}
  \caption{Pressure data alignment}
  \label{fig_pressure_align}
\end{figure}

Due to imperfections in hardware design, the pressure insoles for the left and right feet, which transmit data to a computer via Wi-Fi, experience asynchronous arrival times at the computer end. As depicted in Fig.~\ref{fig_align:sub1}, there is a noticeable time offset between the data from the left and right foot insoles. To synchronize the data from both pressure insoles, users are instructed to perform an action of lifting the heel and then placing the sole on the ground before starting data collection, maintaining this position for 3 seconds. This specific action serves as a characteristic feature to align the data from both feet. We calculated the $Y_{\text{CoP}} $ of both feet using Equation \ref{equ:cop} and used this parameter to identify the above features and align the data. 

\begin{equation}
    Y_{\text{CoP}} = \frac{\sum_{i=1}^{n} P_i \cdot y_i}{\sum_{i=1}^{n} P_i}
    \label{equ:cop}
\end{equation}
where \( P_i \) is the pressure reading from sensor \( i \), and \( y_i \) is the y-coordinate of sensor \( i \), for \( i = 1 \) to \( n \), where \( n \) is the total number of sensors.

\begin{figure}[htbp]
    \centering
    \includegraphics[width=0.7\linewidth]{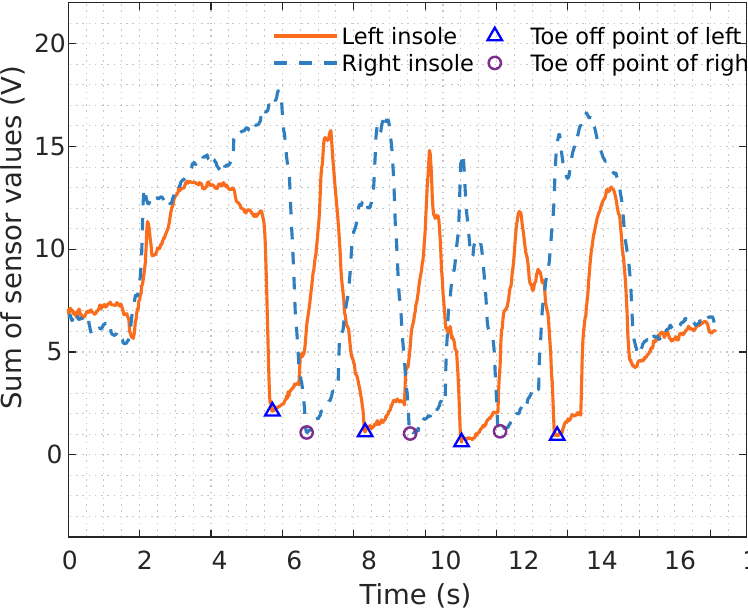} 
    \caption{Move first detection.} 
    \label{fig:move first} 
\end{figure}
The synchronized data, post-alignment, is illustrated in Fig.~\ref{fig_align:sub2}, where the CoP's y-axis coordinate effectively aligns the timings of the pressure data from both insoles.


\subsubsection{Walking Time and Speed Calculation}
\begin{figure}[htbp]
  \centering
  \begin{subfigure}[b]{0.47\linewidth} 
    \includegraphics[width=\linewidth]{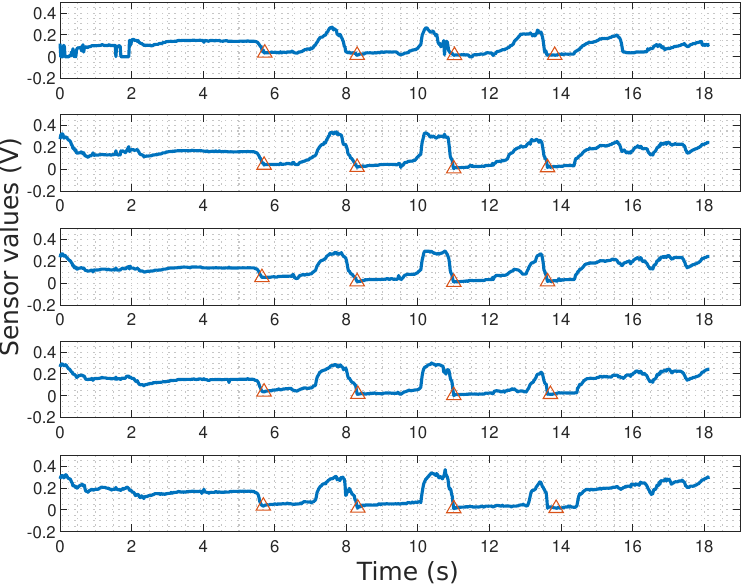}
    \caption{the raw data}
    \label{fig_time:sub1}
  \end{subfigure}%
  \hspace{2mm} 
  \begin{subfigure}[b]{0.46\linewidth} 
    \includegraphics[width=\linewidth]{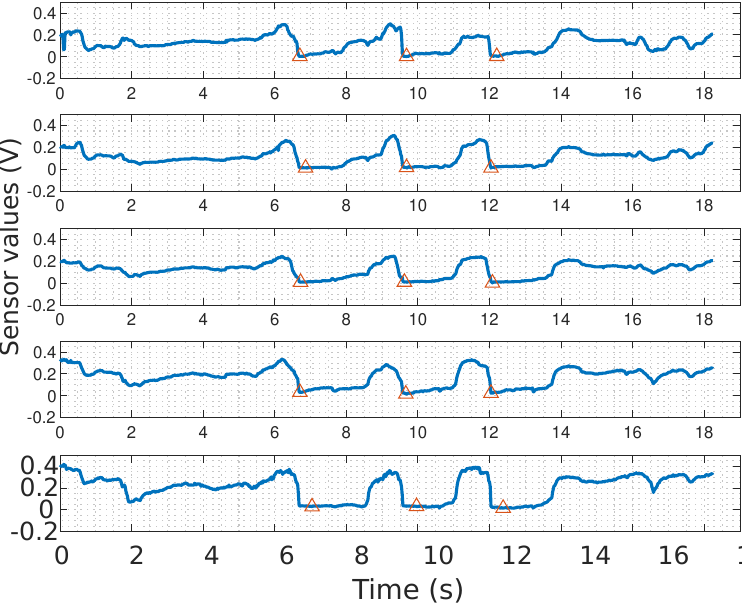}
    \caption{alignment data}
    \label{fig_time:sub2}
  \end{subfigure}
  \caption{Walking time calculation.}
  \label{fig_pressure_time}
\end{figure}
We define the start of walking as the moment the toe of one foot leaves the ground, and the end of walking as the instant the toe of the other foot lifts off. Based on empirical thresholds, we utilize the sum of sensor values to determine which foot moved first. Additionally, the five sensor values from row 1 of each insole are used to calculate the walking time, as illustrated in Fig. \ref{fig:move first} and Fig. \ref{fig_pressure_time}, respectively. The walking velocity is calculated from the relation between the walking time and the walking velocity.

\subsection{CSI-pressure Fusion Model}
\subsubsection{Geometric Relationship}
\begin{figure}[H]
    \centering
    \includegraphics[width=0.7\linewidth]{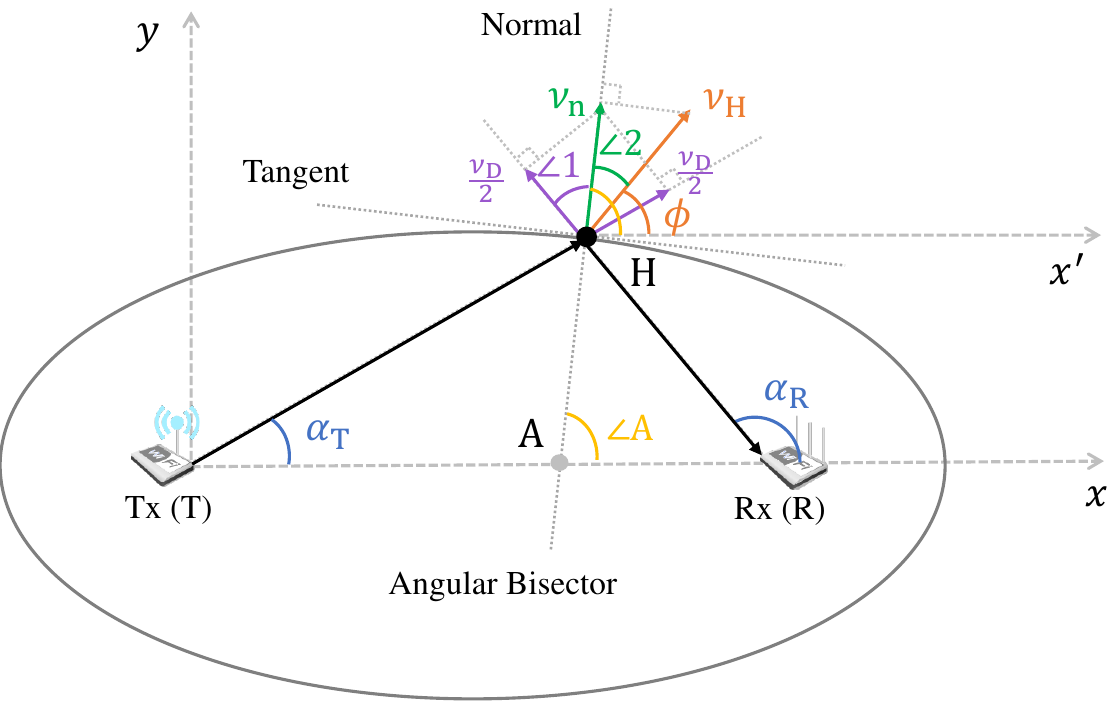} 
    \caption{The relationship between walking velocity and Doppler velocity.} 
    \label{fig:fresnel_zone} 
\end{figure}

Focusing on \( \text{Tx} \) and \( \text{Rx} \) as the centers, concentric ellipses form Fresnel zones in the monitoring space. As a person walks, they cross several boundaries of these Fresnel zones. A Doppler effect arises when the target has a non-zero velocity in the normal direction of the ellipse. However, if the target consistently moves along the tangent of the ellipse, no Doppler effect is induced. We define the speed of path length variation computed from the Doppler frequency shift as the Doppler velocity, \( v_D \). The relationship between a person's walking speed, \( v_H \), walking direction, \( \phi \), and the Doppler velocity is illustrated in the Fig. ~\ref{fig:fresnel_zone}. 
In Fig. \ref{fig:fresnel_zone}, \( \text{Tx} \) represents the signal transmission device, \( \text{Rx} \) denotes the signal reception device, \( \alpha_T \) is the angle of signal departure, \( \alpha_R \) is the angle of signal arrival, and point \( H \) indicates the current position of the moving target. \( v_n \) is the velocity component of \( v_H \) in the normal direction of the ellipse at point \( H \). The normal containing \( v_n \) intersects the x-axis at point \( A \). Given the properties of the ellipse, the angle bisector of \( \angle \text{THR} \) coincides with the normal. The velocity \( v_n \) is decomposed into \( v_D/2 \) along the extensions of \( \text{TH} \) and \( \text{RH} \). Based on the above relationships, we can derive the following geometric relations:

\begin{figure}[H]
    \centering
    \includegraphics[width=0.83\linewidth]{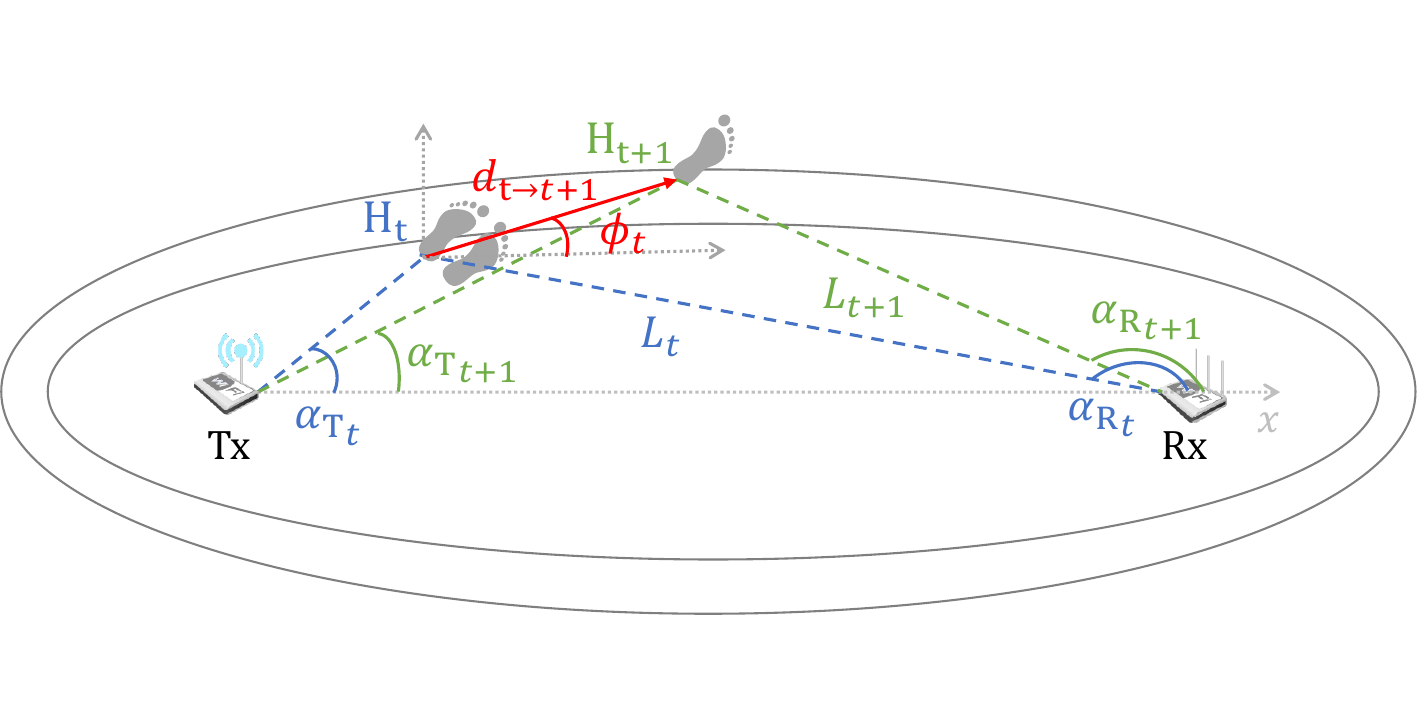} 
    \caption{The relationship between stride length and reflection path length.} 
    \label{fig:reflected_path} 
\end{figure}

\begin{align}
\left\{ \begin{aligned}
\angle 1 & = \frac{\alpha_T - \alpha_R}{2} \\
\angle 2 & = \frac{\alpha_T + \alpha_R}{2} - \phi \\
v_D      & = 2v_n \cos \angle 1 \\
v_n      & = v_H \cos \angle 2
\end{aligned} \right.
\label{eq:vd_vh}
\end{align}

By simplifying Equation \ref{eq:vd_vh}, the relationship between the Doppler velocity \( v_{D_t} \) and the walking speed \( v_{H_t} \) at time \( t \) can be derived as:
\begin{equation}
2\cos\left(\frac{\alpha_{T_t} - \alpha_{R_t}}{2}\right)\cos\left(\frac{\alpha_{T_t} + \alpha_{R_t}}{2} - \phi_t\right) = \frac{v_{D_t}}{v_{H_t}}
\label{eq:model_1}
\end{equation}


\subsubsection{Reflected Signal Path Length Variation Model}

When an individual walks within a monitoring area equipped with Wi-Fi devices while wearing pressure insoles, the insoles record the time taken for each step of the walker, while the CSI signal captures the variations in the reflected signal path length caused by the walking process. Assuming that a person's stride remains consistent in an indoor environment, we can model the reflected path length provided by CSI with the stride length and time information offered by the insoles. As illustrated in Fig. \ref{fig:reflected_path}, an individual moves from the position \( H_{t} \) to \( H_{t+1} \) with a step, covering a distance \( d \) in the direction \( \phi_{t} \). \( L_{t} \) and \( L_{t+1} \) represent the reflected path lengths at times \( t \) and \( t+1 \), respectively. Using these parameters, we can establish a mathematical relationship between the reflected signal path length, stride length and walking direction:

\begin{equation}
L_{t+1} - L_t = \int_{t}^{t+1} v_D(\tau) \, d\tau
\label{eq:reflected_path}
\end{equation}

We define the coordinates of \( \text{Tx} \) and \( \text{Rx} \) as vectors \( \vec{l}_{T} \) and \( \vec{l}_{R} \), respectively. The coordinates of the target at \( H_{t} \) and \( H_{t+1} \) are represented by vectors \( \vec{l}_{H_t} \) and \( \vec{l}_{H_{t+1}} \), respectively. \( L_{t} \) and \( L_{t+1} \) can be expressed as:

\begin{equation}
\small
\left\{
\begin{array}{l}
L_t = \left\| \vec{l}_{H_t} - \vec{l}_T \right\| + \left\| \vec{l}_{H_t} - \vec{l}_R \right\| \\[1em]
L_{t+1} = \left\| \vec{l}_{H_{t+1}} - \vec{l}_T \right\| + \left\| \vec{l}_{H_{t+1}} - \vec{l}_R \right\|
\end{array}
\right.
\label{eq:length}
\end{equation}

The relationship between the coordinates of \( \vec{l}_{H_t} \) and \( \vec{l}_{H_{t+1}} \) can be expressed as:\\

\begin{equation}
\left\{
\begin{array}{l}
x_{t+1} = x_t + d \cos \phi_t \\
y_{t+1} = y_t + d \sin \phi_t
\end{array}
\right.
\label{eq:coor_re}
\end{equation}

\subsubsection{CSI-pressure Fusion Model}


By combining Equations \ref{eq:model_1} and \ref{eq:reflected_path}, we integrated walking information from Wi-Fi and pressure insoles to establish two geometric models. In both equations, the right-hand side is known, while the left-hand side parameters—coordinates and walking direction \((x, y, \phi)\)—are to be estimated. These equations serve as constraints. 
By utilizing the geometric intersection of an ellipse and a line with the person's position, we construct an objective function to find the optimal solution, defined as follows:
\begin{equation}
(\hat{x}_t, \hat{y}_t, \hat{\phi}_t) = \underset{(x_t, y_t, \phi_t)}{\text{arg min}} \, |f(x_t, y_t, \phi_t)|
\end{equation}


\begin{equation}
\begin{split}
f(x_t, y_t, \phi_t) = & \, \underbrace{\left| \frac{(x_t - d_{LoS}/2)^2}{a_t^2} + \frac{y_t^2}{b_t^2} - 1 \right|}_{\circled{1}} \\
& \quad + \underbrace{|y_t - \tan\alpha_{T_t} \cdot x_t|}_{\circled{2}}
\end{split}
\label{eq:model_3}
\end{equation}




The term {\circled{1}} denotes the target's elliptical equation, while {\circled{2}} indicates its linear equation.They intersect at the position of the target. In the ellipse equation, \( a_t \) and \( b_t \) are defined as:

\begin{equation}
\left\{
\begin{aligned}
a_t &= \frac{L_t}{2} \\
b_t &= \sqrt{a_t^2 - c^2} \\
c &= \frac{\| \vec{l}_R - \vec{l}_T \|_2}{2}
\end{aligned}
\right.
\end{equation}

For the line equation, the slope \( \tan \alpha_{T_t} \) is defined as:
\begin{equation}
\alpha_{T_t} = \alpha_{T_{t+1}} + \beta
\end{equation}
where \( \beta \) is the angle between the positions at times \( t \) and \( t+1 \) and the line connecting to the Tx. The definitions of \(\alpha_{T_{t+1}}\) and \(\beta\) are given as:

\begin{align}
\alpha_{T_{t+1}} &= \arctan\frac{y_t + d\sin\phi_t}{x_t + d\cos\phi_t} \\[1em]
\beta &= \arccos\frac{\vec{l}_{H_t} \cdot \vec{l}_{H_{t+1}}}{\|\vec{l}_{H_t}\| \, \|\vec{l}_{H_{t+1}}\|}
\end{align}

\section{Implementation and Evaluation}
\subsection{Implementation}
\begin{figure}[ht]
  \centering
  \begin{subfigure}[b]{0.36\linewidth} 
    \includegraphics[width=\linewidth]{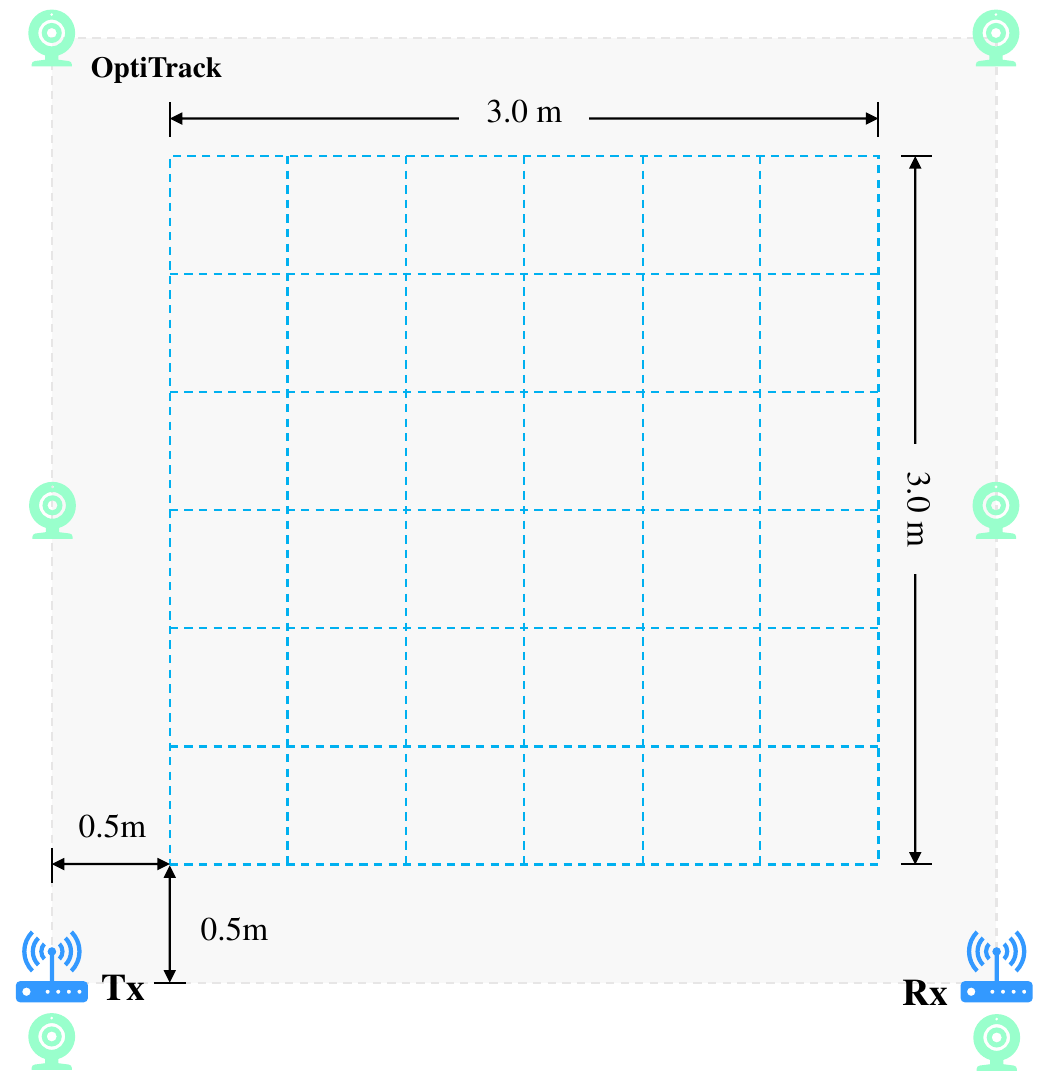}
    \caption{Layout}
    \label{fig_indoor:sub1}
  \end{subfigure}%
  \hspace{2mm} 
  \begin{subfigure}[b]{0.56\linewidth} 
    \includegraphics[width=\linewidth]{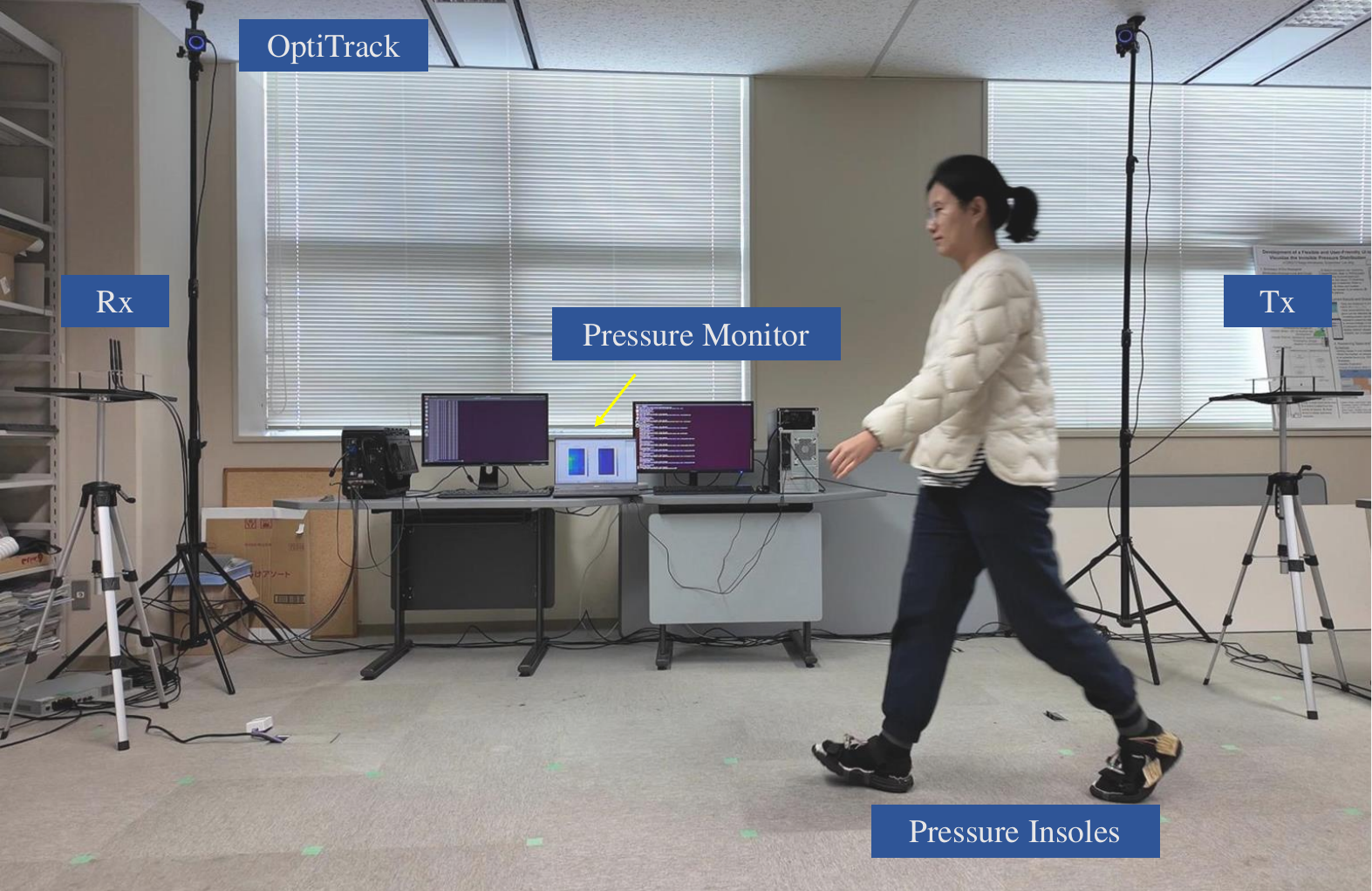}
    \caption{Real indoor environment}
    \label{fig_indoor:sub2}
  \end{subfigure}
  \caption{Experiment environment.}
  \label{fig:indoor layout}
\end{figure}
The experimental setup is depicted in Fig. \ref{fig:indoor layout}, with a sensing area of 4m by 4m. The distance between the Wi-Fi signal transmission and reception devices is 4m, and they are positioned at a height of 1.3m. Participants, wearing pressure insoles, walk within the sensing range, and the pressure data is wirelessly transmitted to a computer via Wi-Fi. We use OptiTrack as the ground truth by placing markers on both feet to capture the walker's position. The Wi-Fi equipment has a sampling rate of 1000Hz, the pressure insoles have a sampling rate of 50Hz, and OptiTrack has a sampling rate of 100Hz. Data is processed using MATLAB R2022a.

\subsection{Simulation Experiment}

\begin{figure}
    \centering
    \includegraphics[width=0.65\linewidth]{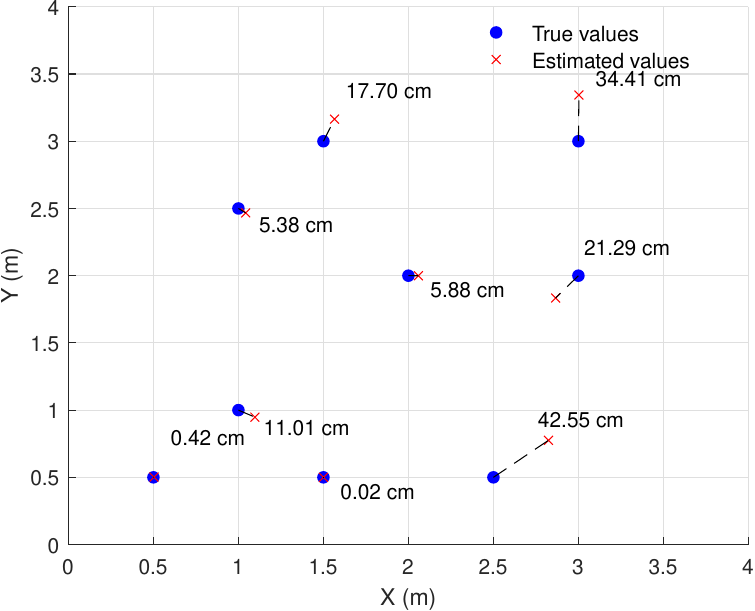} 
    \caption{Initial position estimation.}
    \label{fig:simulation_results} 
\end{figure}

We perform a simulation validation of the CSI-pressure fusion model to test the model's performance in estimating the initial position. The simulation results are shown in Fig .\ref{fig:simulation_results}. In the figure, the x and y axes represent the spatial coordinates in meters. The blue dots indicate the ground truth values, while the red markers represent the estimated values obtained using the model. The dashed lines between the ground truth and the estimated values indicate the error. From the figure, it can be seen that the estimation error for the initial position ranges from 0.02 cm to 42.55 cm. In addition, the average of initial positioning errors is 15.41 cm, while the unbiased standard deviation is 15.04 cm. 

\subsection{Trajectory Estimation}
\begin{figure}
    \centering
    \includegraphics[width=0.7\linewidth]{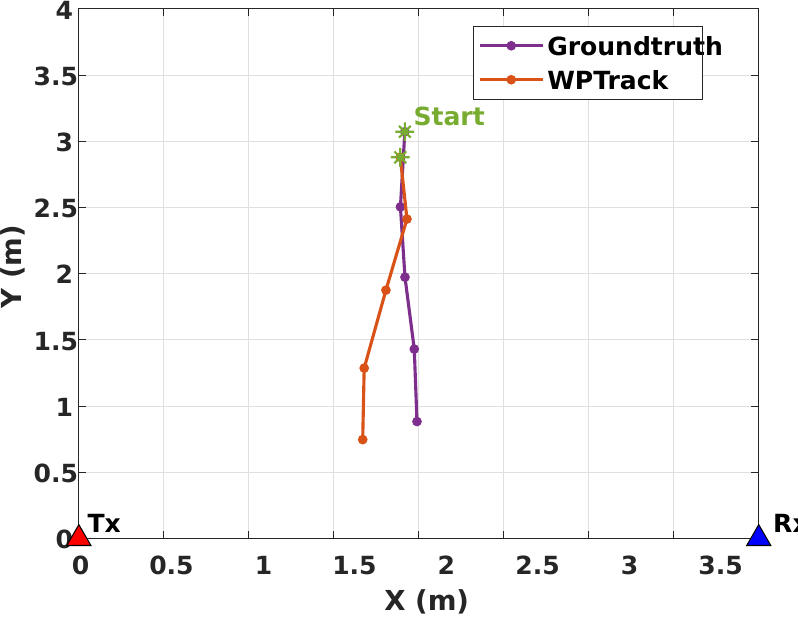} 
    \caption{Trajectory estimation result.}
    \label{fig:trajectory_estimation} 
\end{figure}
In a real-world scenario, we have volunteers wearing pressure insoles walking within a Wi-Fi sensing area. By acquiring CSI and plantar pressure data, we estimate the starting point of the walking trajectory and track the target. Fig. \ref{fig:trajectory_estimation} shows the trajectory estimation results for walking along a straight line. The x and y axes represent spatial coordinates in meters, with red and blue triangles representing Tx and Rx, respectively. The asterisks denote the starting points of the trajectories, with purple indicating the ground truth and orange indicating the WPTrack estimated results. As observed from the figure, the WPTrack estimation of the starting point is very close to the ground truth. The trajectory consists of four steps, and it can be seen that the estimated results closely match the actual measurements.

\section{Conclusion}
In this paper, we introduce WPTrack, the first system to fuse Wi-Fi with pressure insoles for single-target tracking. We propose a CSI-pressure fusion model that integrates CSI with pressure data, addressing the challenges of tracking blind spots and initial position estimation inherent to single Wi-Fi link systems. Currently, WPTrack has completed the model development and simulation stages. Moving forward, we plan to conduct real-world tests with five volunteers to validate the system's performance in practical scenarios.


\bibliographystyle{IEEEtran}

\bibliography{references}      

\end{document}